# Effect of deoxygenation on the weak – link behavior of $YBa_2Cu_3O_{7-\delta}$ superconductors

H. Salamati P. Kameli*

Department of Physics, Isfahan University of Technology, Isfahan 84154, Iran

**Abstract**

A systematic study of the weak- link behavior for $YBa_2Cu_3O_{7-\delta}$ polycrystalline samples has been done using the electrical resistivity and AC susceptibility techniques. The experiments were performed with two samples of similar grain, a sample of well - coupled grains, and a deoxygenated sample in such a way that the oxygen mostly come from the intergrain region. Analysis of the temperature dependence of the AC susceptibility near the transition temperature (Tc) has been done employing Bean's critical state model. The observed variation of intergranular critical current densities(Jc) with temperature indicates that the weak links are changed from superconductor- normal metal- superconductor (SNS) for well- coupled sample to superconductor- insulator- normal metal superconductor(SINS) type of junctions for the deoxygenated sample. These results are interpreted in terms of oxygen depletion from grain boundaries, which in turn decreases the intergranular Josephson coupling energy with a consequent decrease of pinning of the intergranular vortices.

---

*Corresponding author. Tel: +98 3113912375, fax: +98 3113912376,
e-mail: pkameli@sepahan.iut.ac.ir

## 1. Introduction

In the field of electric power applications it is necessary to fabricate superconducting materials with high critical current densities Jc. It has been realized that the polycrystalline superconductors can be described as arrays of superconducting grains weakly coupled by Josephson junctions. These weakly coupled grains are known to limit the Jc values of superconductors[1-2].The possible reason for the formation of these weak links are misorientation of grain boundaries and composition variations at the grain boundaries[3-4]. But the details of the current limiting mechanisms being not well understood. In order to see the effect of oxygen deficiency on the intergranular coupling of high-Tc superconductors, in this paper we study the weak link behavior in a deoxygenated polycrystalline samples of the $YBa_2Cu_3O_{7-\delta}$ (YBCO) superconductors in such a way that the oxygen losses mostly come from the intergrain region, while the grains remain essentially unchanged.

## 2. Experimental

YBCO samples were prepared by a conventional solid-state reaction method. High purity powders of $Y_2O_3$, $BaCo_3$, and CuO were well mixed in stoichiometric proportions, calcined at 930 $^{o}$c for 24 h and sintered at 950 $^{o}$c in an oxygen atmosphere for 24 h. The calcinations and grinding procedures were repeated three times. Bars of sintered ceramics (approximate dimensions of $10\times1.5\times1.5mm^3$) were cut from the pellets. The sample in as-sintered conditions will be referred as A. one of the samples was deoxygenated. For this aim it was introduced for 10 min in a long tubular furnace, previously stabilized at 450 $^{o}$c in air, and then quenched to room temperature. This sample denoted by B. The resistivity measurements were carried out by the four probe method. The AC susceptibility measurements were performed using a Lake Shore AC Susceptometer Model 7000. X-ray diffraction (XRD) patterns of samples were taken on a Philips XPERT x-ray difractometer. The microstructure and energy dispersive x-ray (EDX) spectra were taken on Philips XL 30 scanning electron microscope.

**3. Results and discussion**

It is known that the High-Tc granular superconductors having a well defined superconducting transitin temperature, generaly display a two step resistive transition ρ(T) and correspondingly the derivative of the ρ(T) displays a peak and a tail in the lower temperature side [5]. The peak temperature marks the superconducting transition whitin the ragins and the tail is related to the intergranular coupling.

Fig. 1(a) shows the resistive transition behavior of the samples A and B. It can be seen from these curves that, the normal state resistivity regularly increases and the tail regime clearly visible in the ρ(T) curve of the sample B. But the sample A has an almost sharp transiton ($\Delta T \approx 1$ K). The transition temperature taken as the maximum of the derivative of the ρ(T) curves, is the same for both samples (Fig. 1b) while the zero resistivity temperature is clearly lower for sample B. The increase in the normal state resistivity value for sample B, along with the tail in the resistivity and the constancy in the transition temperature, is the signature that the weak-link network is affected by the short time annealing in air, while the grains remain unchanged. Thus we believe that the feature of the resistivity curve for sample B shows that the oxygen losses mostly come from the intergrain region and consequently reduces the intergranular coupling.

In addition to electrical resistivity, the measurement of AC susceptibility is widely used as a nondestructive method for the determination and characterization of the intergrain component in the polycrystalline high- Tc superconductors [6-9]. In particular the loss component of the AC susceptibility has been used widely to probe the nature of weak links in polycrystalline superconductors. It is also employed to estimate some of the important physical properties like critical current density Jc and effective volume fraction of the superconducting grains [10-13]. Several critical state models have been very successful in accounting for major features of the temperature and field variation of AC susceptibility [14-16].

Experimentally, the real part of AC susceptibility,$\chi'$, in polycrystalline samples shows two drops as the temperature is lowered below onset of diamagnetic transition. The first sharp drop is due to the transition within grains and the second gradual change is due to the occurrence of the superconducting coupling between grains. The imaginary part,$\chi''$, shows a peak which is measure of the dissipation in the sample.

Fig. 2 shows the temperature dependencies of the real,$\chi'$, and imaginary,$\chi''$, parts of AC susceptibility for samples A and B in ac field of 50A/m with a frequency of 333Hz. It can be seen from the curves that, the diamagnetic onset temperature of intrinsic superconducting transition is the same for both samples. But the magnetic transition is much sharper than that of sample B. These results suggest that the intergranular coupling between grains in sample A is better than sample B.

Fig. 3 shows the temperature variation of AC susceptibility curves for various ac fields given in the legend where f =333Hz. It is clear from Fig. 3 that as the filed increases, the peak of $\chi''$shifts to lower temperature and broadens. As we can see the degree of shift for sample B is more than sample A. Also the effect of field on the intergranular component in real part is much more for sample B. The amount of the shift as a function of the field amplitude $H_{ac}$, is proportional to the magnitude or strength of the pinning force. The weaker the pinning the smaller the critical current.

Fig. 4 shows the variation of peak temperature, $T_p$, as a function of ac field amplitude, $H_{ac}$ for samples A and B. As clearly seen from Fig. 4 The rate of shift of peak temperature $T_p$, with ac field amplitude $H_{ac}$, for sample B is higher than sample A. Fig. 5 shows that the applied field amplitude vs. the peak temperature $T_p$, can be fitted power low relation $H_{ac} \sim (1- T_p /T_c)^n$. The best fitting for samples A and B is obtained when the respective values of n is 2 and 1.5. In Fig. 5 although $H_{ac}$ was plotted vs. $T_p$, it can also be related as a $J_c$ vs. temperature. We can estimate the intergranular critical current density as a function of peak temperature $T_p$, by using the Bean critical state model [17]. According to the Bean model, for the bar shaped sample the critical current density at the peak temperature $T_p$, can be written as the formula $J_c \sim H_{ac}/(ab)^{1/2}$ where the cross section of the rectangular bar shaped sample is $2a \times 2b$. Thus Fig. 5 can be considered as a graph of $J_c$ vs. temperature. As showed, our data is in good agreement with $J_c$ dependence of temperature as $J_c \sim (1- T /T_c)^2$ for sample A and $J_c \sim (1- T /T_c)^{1.5}$ for sample B. It is known that the polycrystalline superconductors are arrays of weak links, which behave like Josephson junctions. The temperature dependence of the critical current that flows through the Josephson junctions was calculated. This dependence can be written in the form $J_c \sim (1- T /T_c)^n$. The value of n is 2 for superconductor-normal metal-superconductor (SNS) junctions [18], and 1 for superconductor-insulator-superconductor (SIS) junctions [19]. The values of n between 1 and 2 indicate the formation of superconductor-insulator-normal metal-superconductor (SINS) junctions.

As mentioned above for the sample A, the temperature dependence of $J_c(T)$ is in good agreement with equation $J_c(T) \sim (1- T/T_c)^2$ corresponding to weak links of SNS type. Similarly for sample B the $J_c(T)$ fits equation $J_c(T) \sim (1- T/T_c)^{1.5}$ which is characteristic of SINS weak links.

Comparison of data on the resistivity (Fig.1) and susceptibility behavior (Fig.2, 3) for samples A and B shows that the deoxygenation dose alter the weak-link behavior of the polycrystalline YBCO ceramics. The intergranular coupling of sample B is weakened because of grain boundary oxygen depletion. It is known that the composition variation at the grain boundaries is one of the mechanisms that control the weak link behavior of polycrystalline High- $T_c$ superconductors. Deviations from ideal cation stoichiometry and formation of second phases can be considered as a fundamental mechanism that control electronic transport across grain boundaries. But it was shown that even boundaries with excellent cation composition are weak links [3,20]. On the other hand if there is a defect in the oxygen sub-lattice, it will depress the superconducting order parameter, and if the oxygen concentration is low enough will drive it insulating [21-22]. The presences of layers which are depleted of oxygen or which contain disordered oxygen have been considered in detail [23-24]. Babcock et al [25] showed that less depletion of oxygen at the grain boundary is one reason why some grain boundaries have strong-link characteristics. These considerations support the fact that the oxygen deficiency reduces the intergranular Josephson coupling energy with a consequent decrease of the intergranular critical density and associated pinning of the intergranular vortices.

## 4. Conclusion

We have studied the weak-link behavior of polycrystalline YBCO superconductors with different intergranular coupling using the electrical resistivity and AC susceptibility techniques. The temperature dependence of intergranular critical current densities ($J_c$) indicates that the weak links are changed from SNS (well-coupled sample) to SINS (deoxygenated sample) type. These results are interpreted in terms of oxygen depletion from grain boundaries, which in turn decreases the intergranular Josephson coupling energy with a consequent decrease of pinning of intergranular vortices.

## Acknowledgments

The authors would like to thank Isfahan University of Technology for supporting this project.

**Figure captions**

Fig.1. Temperature dependence of the resistivity (a) and derivative of the resistivity (b) for samples A and B.

Fig. 2. Temperature dependence of the AC susceptibility for samples A and B in ac field amplitude of 50A/m at a frequency of 333Hz.

Fig. 3. Temperature dependence of the AC susceptibility for samples A(a) and B (b) in various ac field amplitudes at a frequency of 333Hz.

Fig. 4. Field amplitude dependence of peak temperature $T_p$ for samples A and B.

Fig. 5. The fitting graphs showing the variation of Field amplitude $H_{ac}$ with $(1- T_p /T_c)^n$ for sample A (n=2) a, and sample B (n=1.5) b.

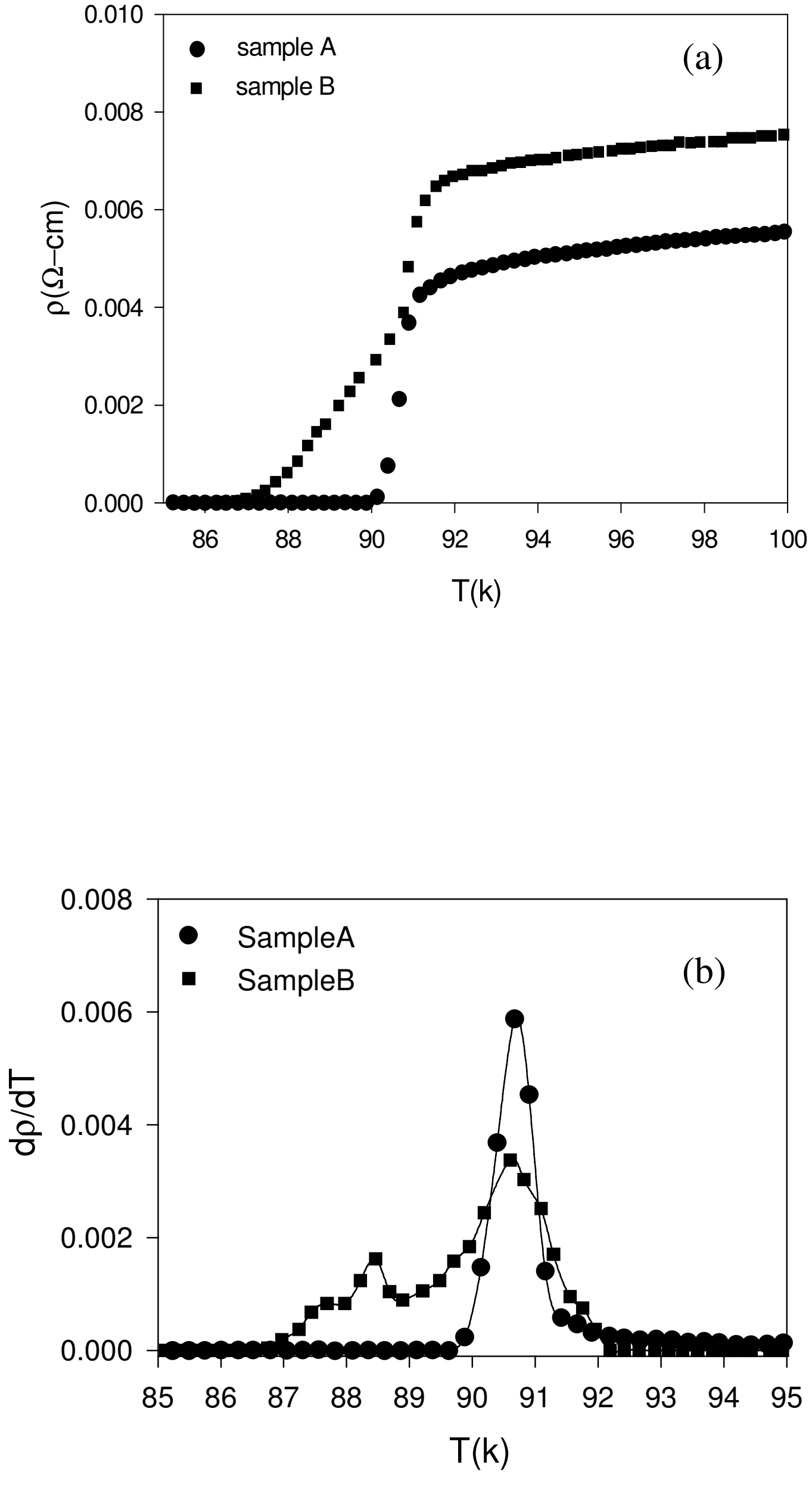

Fig. 1

H. Salamati et al

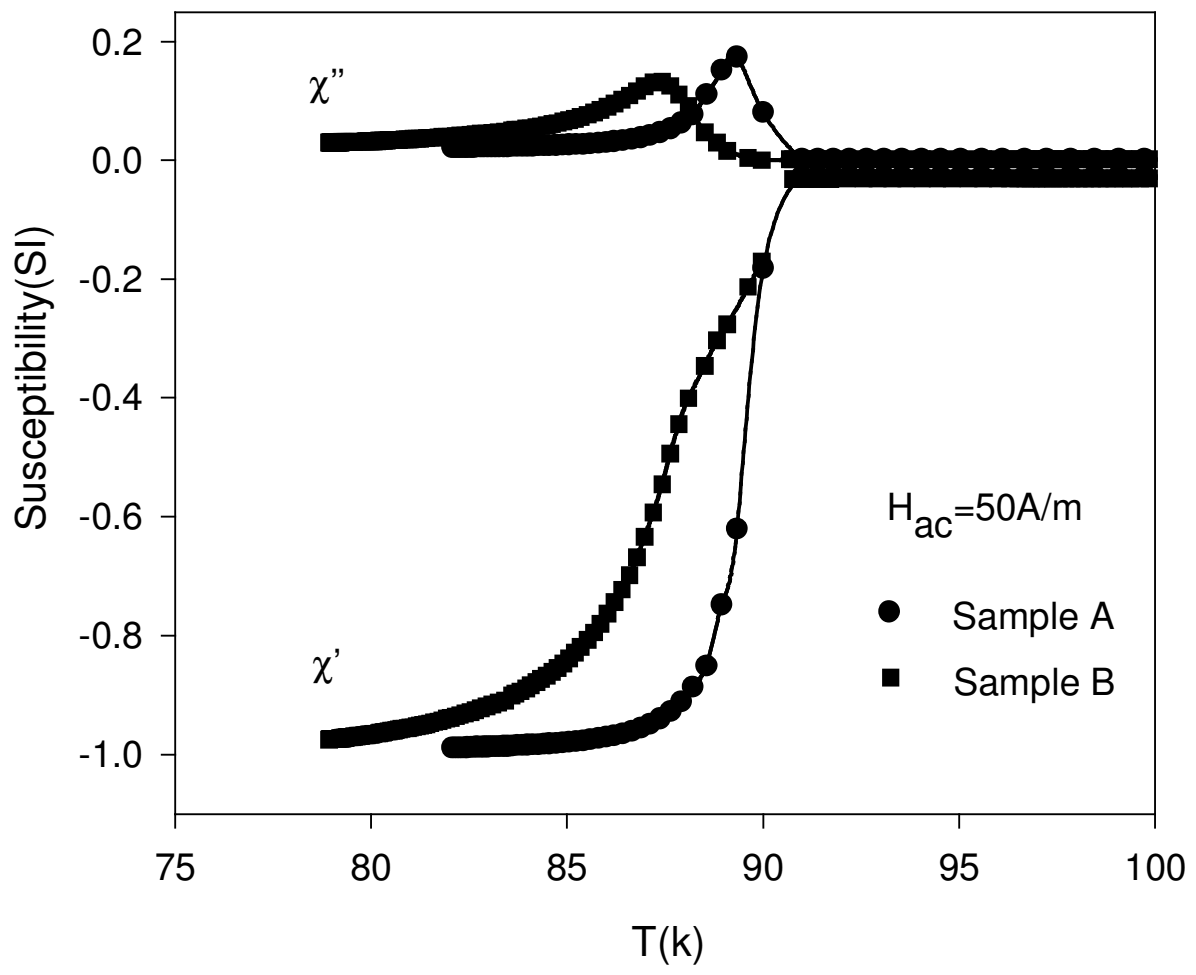

Fig. 2

H. Salamati et al

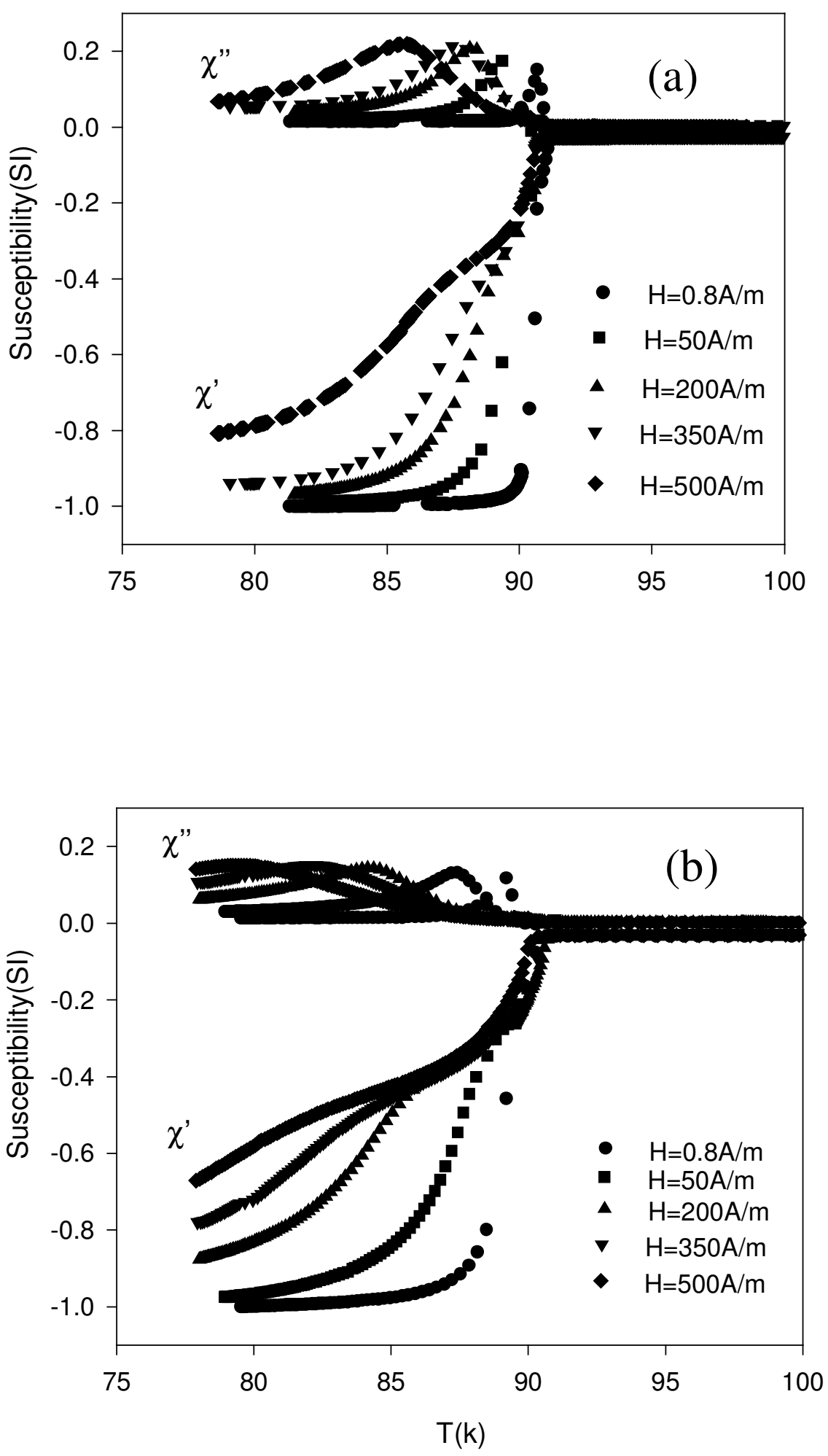

Fig. 3

H. Salamati et al

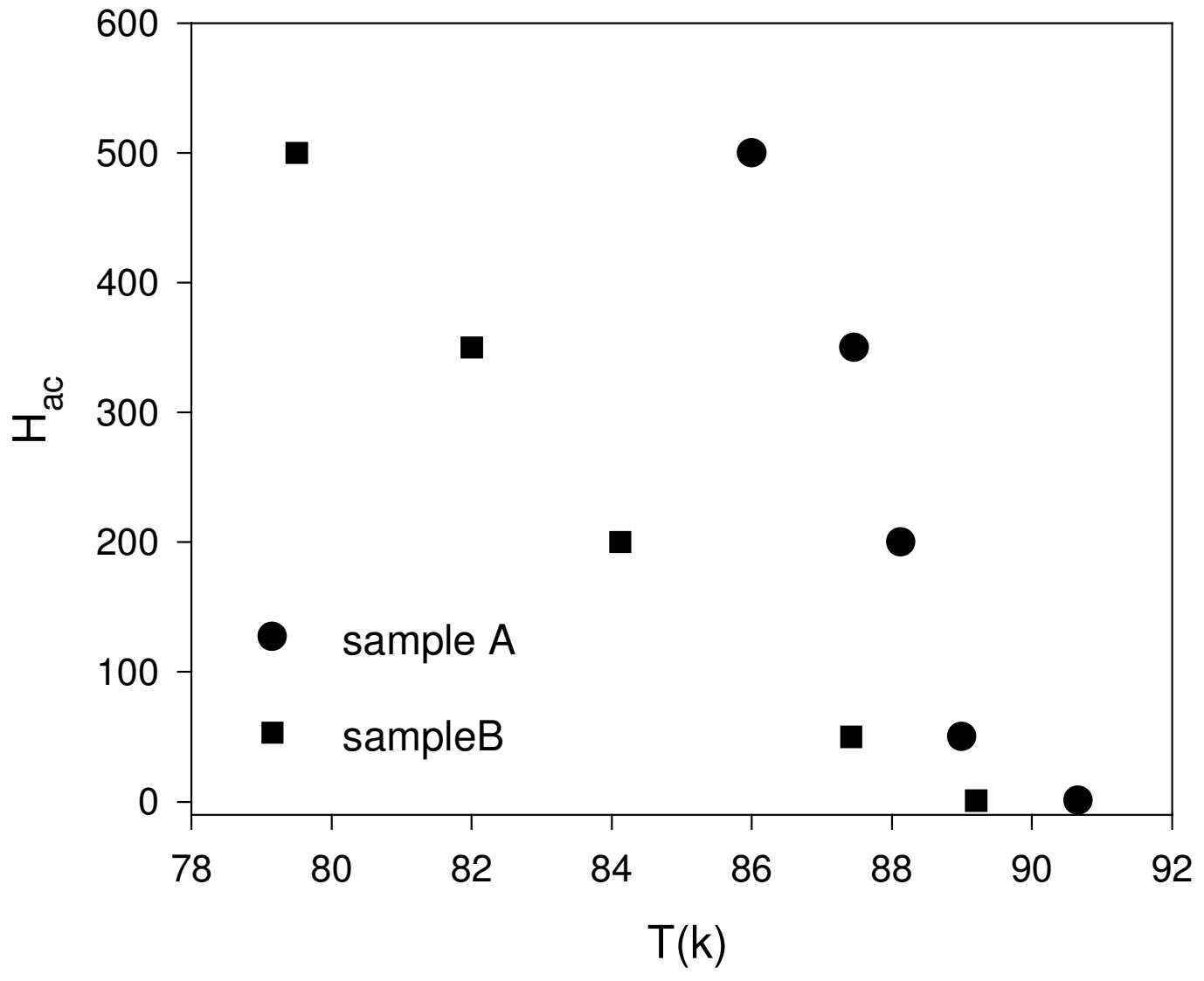

Fig. 4

H. Salamati et al

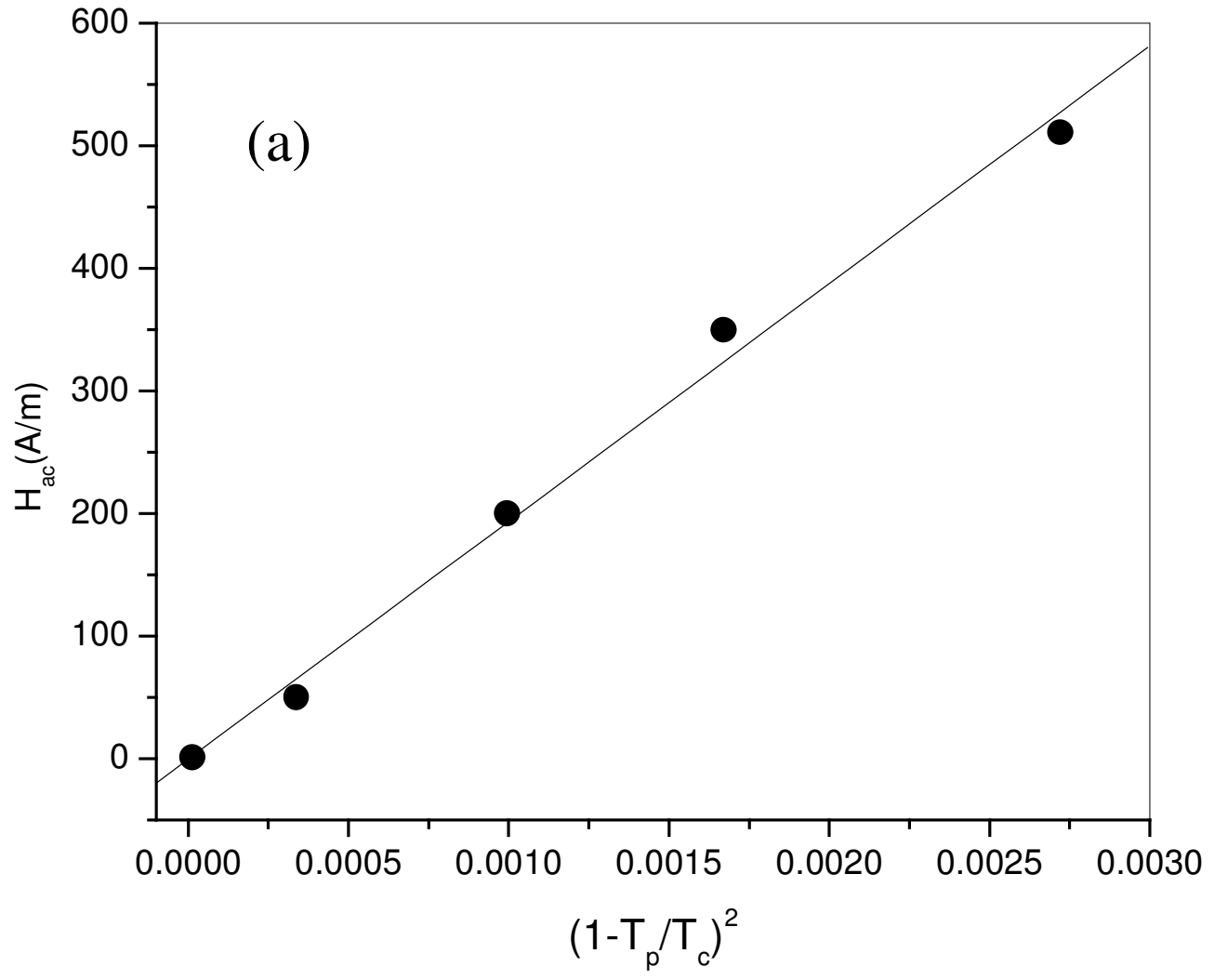

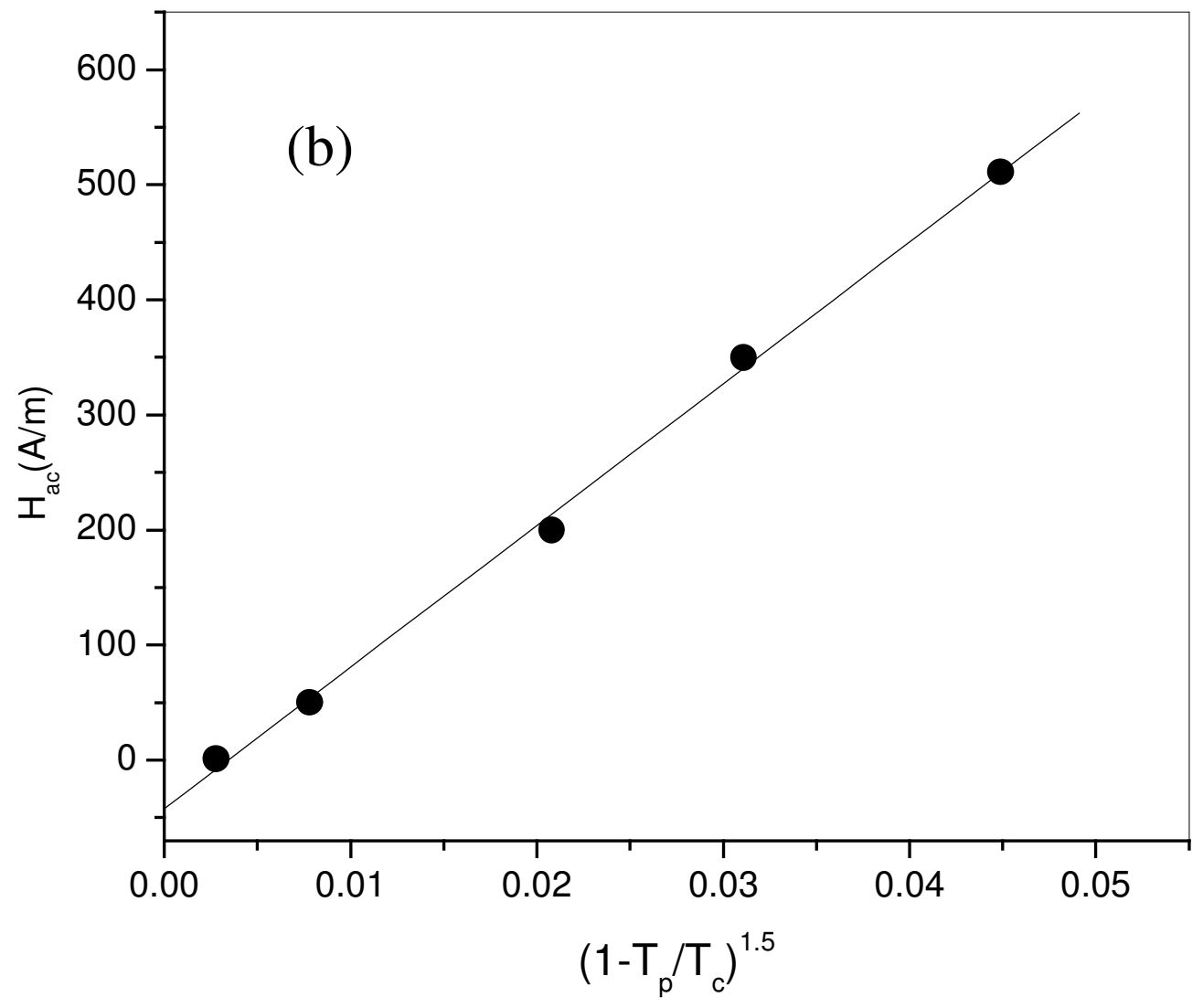

Fig. 5

H. Salamati et al